\documentclass[preprint,eqsecnum,amssymb]{revtex4}

\usepackage{amsmath}

\begin{document}

\title{Entanglement without Dissipation: A Touchstone for an exact Comparison of Entanglement Measures $^{\dagger}$}
\author{G. W. Ford}
\affiliation{Department of Physics, University of Michigan, Ann Arbor, MI 48109-1120 USA}

\author{Yang Gao and R. F. O'Connell}
\affiliation{Department of Physics and Astronomy, Louisiana State University, Baton Rouge, LA  70803-4001 USA}

\date{\today}

\begin{abstract}
Entanglement, which is an essential characteristic of quantum mechanics, is the key element in potential practical quantum information and quantum communication systems.  However, there are many open and fundamental questions (relating to entanglement measures, sudden death, etc.) that require a deeper understanding.  Thus, we are motivated to investigate a simple but non-trivial correlated two-body continuous variable system in the absence of a heat bath, which facilitates an \underline{exact} measure of the entanglement at all times.  In particular, we find that the results obtained from all well-known existing entanglement measures agree with each other but that, in practice, some are more straightforward to use than others. \\
\\
$^{\dagger}$  Dedicated to the memory of Krzysztof Wodkiewicz.
\end{abstract}

\maketitle

\section{Introduction}

Entanglement, which is an essential characteristic of quantum mechanics, plays a key role in all applications related to information science \cite{knight05,amico08,vedral08,horodecki}.  Entanglement describes correlations between two or more particles or subsystems (qubits, oscillators, etc.).  Despite the fact that much insight has already been obtained, much remains to be done as underlined by the title of a recent book on the subject \cite{aczel01}.  For example, whereas there is a plethora of entanglement measures, there is also a consensus that a more unifying fundamental measure needs to be developed.  Also, the discovery of "entanglement sudden death" (ESD) \cite{diosi03,yu04,eberly07}, in contrast to the well-known exponential decay of decoherence, requires a deeper understanding.  Since investigations of ESD have also incorporated heat bath effects, master equations have been the tool of choice, despite their inherent limitations \cite{ford05,ford07}.  Thus, we are motivated to analyze a simple but non-trivial correlated system which displays entanglement in the absence of a heat bath.  For this system, an \underline{exact} measure of entanglement exists which provides us with a touchstone for judging some of the various entanglement measures discussed in the literature.  Since some of these measures involve entropy considerations, this should also throw some light on whether there is a close relationship between the entanglement of a system and its entropy.

Thus, in Sec. II, we will consider a system of two free particles in an initially entangled state amd we calculate its time dependence.  In Sections III, IV and V, we examine the same state using various entanglement criteria and demonstrate explicitely the various steps needed to demonstrate entanglement at all times.  In Sec. VI, we present our conclusions that the various entanglement criteria lead to the same results but that the logarithmic negativity entanglement criterion is the simplest to use.

\section{Entangled Wave Function}

We consider two free particles, each of mass $m$, at positions $x_{1}$ and $x_{2}$, in an initially entangled Gaussian state. Thus, we are dealing with a system with continuous degrees of freedom (as distinct from a system of discrete variables such as qubits), applicable to particle position or momenta or to the field modes of light (of interest in connection with linear optical quantum computing).

The most general initial Gaussian wave function is

\begin{equation}
\psi(x_{1},x_{2};0)=\frac{(a_{11}a_{22}-a^{2}_{12})^{1/4}}{\sqrt{2\pi}}\exp\left\{-\frac{a_{11}x^{2}_{1}+2a_{12}x_{1}x_{2}+a_{22}x^{2}_{2}}{4}\right\}. \label{entm2.1}
\end{equation}
In order that this state be square-integrable we must of course assume that $a_{11}$ and $a_{22}$ are positive and that $a_{11}a_{22}-a^{2}_{12}>0$.  We specialize to the symmetric case, by choosing

\begin{equation}
a_{11}=a_{22}=\frac{1}{\sigma^{2}}+\frac{1}{4d^{2}},~~~a_{12}=\left(-\frac{1}{\sigma^{2}}+\frac{1}{4d^{2}}\right)<0, \label{entm2.2}
\end{equation}
so that $2d>\sigma$ (the opposite choice would lead to similar conclusions).  As we shall see presently, $d$ corresponds to the width of the center-of -mass system. Thus, $a_{12}$ provides a measure of entanglement at $t=0$. In fact, since we are dealing with free particles we expect that the entanglement will not change in time.  However, as we shall see, the time-dependent coefficient of $x_{1}x_{2}$ will \underline{not} serve as a measure of entanglement.

The above equations enable us to write (\ref{entm2.1}) in the form

\begin{equation}
\psi(x_{1},x_{2};0)=\frac{1}{\sqrt{2\pi\sigma d}}\exp\left\{-\frac{(x_{1}-x_{2})^{2}}{4\sigma^{2}}-\frac{(x_{1}+x_{2})^{2}}{16d^{2}}\right\}. \label{entm2.3}
\end{equation}
Next, we transform to center of mass and relative coordinates \cite{merzbacher98}

\begin{equation}
x=x_{1}-x_{2}; ~~~ X=(x_{1}+x_{2})/2 \label{entm2.4}
\end{equation}

\begin{equation}
m_{x}=m/2;~~~ M=2m \label{entm2.5}
\end{equation}
to give the result

\begin{equation}
\psi(x_{1},x_{2};0)=\phi(x)\Psi(X)=\frac{1}{\sqrt{2\pi\sigma d}}\exp\left\{-\frac{x^{2}}{4\sigma^{2}}-\frac{X^{2}}{4d^{2}}\right\}. \label{entm2.6}
\end{equation}
It is clear that $\sigma$ is the width of the relative coordinate system whereas $d$ is the width of the center-of-mass coordinate system.

For future reference, we also note that the corresponding relative and center of mass momenta are

\begin{equation}
p=\frac{1}{2}(p_{1}-p_{2}); ~~~ P=p_{1}+p_{2}, \label{entm2.7}
\end{equation}
respectively.  It is clear that both $(x,p)$ and $(X,P)$ satisfy the usual commutation relations for conjugate canonical variables.

Since our wave function in the transformed coordinates now behaves as the product of two independent Gaussian wave packets, we can now apply the exact propagation method for a free particle \cite{merzbacher98} to obtain the wave function at a time $t$, with the result

\begin{eqnarray}
&&\hspace{-2cm}\psi(x_{1},x_{2};t) = \sqrt{\frac{2\sigma d}{\pi\left(\sigma^{2}+\frac{i\hbar t}{m}\right)\left(4d^{2}+\frac{i\hbar t}{m}\right)}}\exp\left\{-\frac{(x_{1}-x_{2})^{2}}{4(\sigma^{2}+\frac{i\hbar t}{m})}-\frac{(x_{1}+x_{2})^{2}}{4(4d^{2}+\frac{i\hbar t}{m})}\right\} \nonumber \\
&&\hspace{-0.04cm}= \sqrt{\frac{2\sigma d}{\pi(\sigma^{2}+\frac{i\hbar t}{m})(4d^{2}+\frac{i\hbar t}{m})}}\exp\left\{-\frac{x^{2}}{4(\sigma^{2}+\frac{i\hbar t}{m})}-\frac{X^{2}}{4(d^{2}+\frac{i\hbar t}{4m})}\right\} \nonumber \\
&&\hspace{-0.04cm}= \sqrt{\frac{2\sigma d}{\pi(\sigma^{2}+\frac{i\hbar t}{m})(4d^{2}+\frac{i\hbar t}{m})}}\exp\left\{-\frac{x^{2}}{4\sigma^{2}(t)}\left[1-\frac{i\hbar t}{m\sigma^{2}}\right]-\frac{X^{2}}{4\sigma^{2}_{d}(t)}\left[1-\frac{i\hbar t}{4md^{2}}\right]\right\} , \label{entm2.8}
\end{eqnarray}
where

\begin{equation}
\sigma^{2}(t)=\sigma^{2}\left\{1+\left(\frac{\hbar t}{m\sigma^{2}}\right)^{2}\right\}, \label{entm2.9}
\end{equation}
and

\begin{equation}
\sigma^{2}_{d}(t)=d^{2}\left\{1+\left(\frac{\hbar t}{4md^{2}}\right)^{2}\right\}. \label{entm2.10}
\end{equation}
It follows that \cite{ford02}

\begin{equation}
\langle x(t) \rangle=0;~~ \langle X(t)\rangle =0;~~ \langle p(t)\rangle =0;~~ \langle P(t)\rangle =0 \label{entm2.11}
\end{equation}

\begin{equation}
\langle x^{2}(t) \rangle =\sigma^{2}(t);~~~ \langle X^{2}(t)\rangle=\sigma^{2}_{d}(t) \label{entm2.12}
\end{equation}
and

\begin{equation}
\langle p^{2}(t)\rangle =\frac{\hbar^{2}}{4\sigma^{2}};~~~ \langle P^{2}(t)\rangle =\frac{\hbar^{2}}{4d^{2}}. \label{entm2.13}
\end{equation}
We also note that

\begin{equation}
a_{11}(t)=a_{22}(t)=\frac{1}{\sigma^{2}(t)}+\frac{1}{4\sigma^{2}_{d}(t)} . \label{entm2.14}
\end{equation}
In addition, we denote the coefficient of the real part of the $\left(-\frac{1}{2} x_{1}x_{2}\right)$ power in the exponential in (\ref{entm2.8}) by $a_{12}(t)$ to obtain

\begin{equation}
a_{12}(t)=\left\{-\frac{1}{\sigma^{2}(t)}+\frac{1}{4\sigma^{2}_{d}(t)}\right\}. \label{entm2.15}
\end{equation}
However, it is now clear that $a_{12}(t)$ can \underline{not} be used as a measure of entanglement because it predicts that entanglement decreases in time, eventually falling to zero [see the following section, especially (\ref{entm3.16})] whereas we expect that entanglement does not change in time.  On the other hand, we have the tools to investigate the results obtained by use of various entanglement measures discussed in the literature.

More generally, we point out that the starting point for all investigations is the well-known separability condition

\begin{equation}
\rho=\sum_{j}c_{j}\rho_{j}^{(1)}\otimes\rho^{(2)}_{j}, \label{entm2.16}
\end{equation}
where $\rho$ is the density matrix of the quantum state which is written as a convex combination of tensor product states for the $j$ states and where the individual terms are normalized so that $\sum_{j}c_{j}=1$.  If this decomposition is not possible, then we say that the state is entangled. We now turn to some specific entanglement criteria which have been proposed.

\section{Duan et al. \cite{duan00} Criterion}

Using the uncertainty principle, Duan et al. \cite{duan00} derived a sufficient criterion for inseparability for a pair of EPR type operators for continuous variable systems.  Further work on this topic appears in \cite{mancini02,nha08}.  In particular, this approach applies to our problem.  Following \cite{duan00}, we write 

\begin{equation}
u=\frac{1}{L}\left(|a|x_{1}+\frac{1}{a}x_{2}\right),~~~v=\frac{L}{\hbar}\left(|a|p_{1}-\frac{1}{a}p_{2}\right). \label{entm3.1}
\end{equation}
(except that we have introduced the parameter $L$ which has the dimension of length), where $a$ is an arbitrary, non-zero real number.  Use of the uncertainty relation \cite{duan00} leads to the result

\begin{equation}
\frac{1}{L^{2}}\left\langle \Delta u^{2}\right\rangle +\frac{L^{2}}{\hbar^{2}}\left\langle \Delta v^{2}\right\rangle \geqslant\left(a^{2}+\frac{1}{a^{2}}\right), \label{entm3.2}
\end{equation}
for any $L$. For $L=1$, this is the Duan et al. result.  However, by minimizing with respect to $a$ and $L$, we will obtain an improved version given by (\ref{entm3.11}) below.

For a state that is \emph{not} entangled, form

\begin{equation}
\langle u^{2}\rangle +\langle v^{2}\rangle =\frac{a^{2}}{L^{2}}\langle x^{2}_{1}\rangle +\frac{L^{2}a^{2}}{\hbar^{2}}\langle p^{2}_{1}\rangle+\frac{1}{L^{2}a^{2}}\langle x^{2}_{2}\rangle+\frac{L^{2}}{a^{2}\hbar^{2}}\langle p^{2}_{2}\rangle . \label{entm3.3}
\end{equation}
Here we have used the fact that for a non-entangled state (\ref{entm2.16}) the quantities

\begin{eqnarray}
\langle x_{1}x_{2}\rangle &=& \langle x_{1}\rangle\langle x_{2}\rangle =0, \label{entm3.4} \\
\langle p_{1}p_{2}\rangle &=& \langle p_{1}\rangle\langle p_{2}\rangle =0, \label{entm3.5}
\end{eqnarray}
because we have restricted our discussion to states for which

\begin{equation}
\langle x_{1}\rangle =\langle x_{2}\rangle =\langle p_{1}\rangle =\langle p_{2}\rangle =0. \label{entm3.6}
\end{equation}
(Duan et al. do not make this restriction but come to the same conclusion for what they call $\Delta u$ and $\Delta v$.)  Next we use the uncertainty principle,

\begin{equation}
\langle p^{2}_{1}\rangle \geq\frac{\hbar^{2}}{4\langle x^{2}_{1}\rangle},~~~ \langle p^{2}_{2}\rangle\geq \frac{\hbar^{2}}{4\langle x^{2}_{2}\rangle}, \label{entm3.7}
\end{equation}
to get

\begin{equation}
\langle u^{2}\rangle +\langle v^{2}\rangle\geq\left(\frac{\langle x^{2}_{1}\rangle}{L^{2}}+\frac{L^{2}}{4\langle x^{2}_{1}\rangle}\right)a^{2}+\left(\frac{\langle x^{2}_{2}\rangle}{L^{2}}+\frac{L^{2}}{4\langle x^{2}_{2}\rangle}\right)\frac{1}{a^{2}}. \label{entm3.8} 
\end{equation}
Now the quantity $y+\frac{1}{4y}$, $0\leq y <\infty$ has a minimum value of 1 at $y=\frac{1}{2}$. We conclude

\begin{equation}
\langle u^{2}\rangle +\langle v^{2}\rangle\geq a^{2}+\frac{1}{a^{2}}, \label{entm3.9}
\end{equation}
independent of $L$.  This is the result of Duan et al.  Since the inequality is independent of $L$, we can minimize the left hand side with respect to $L$ to get

\begin{equation}
2\sqrt{\left\langle \left(\vert a\vert x_{1}+\frac{1}{a}x_{2}\right)^{2}\right\rangle\left\langle\left(\vert a\vert p_{1}-\frac{1}{a}p_{2}\right)^{2}\right\rangle}\geq\left(a^{2}+\frac{1}{a^{2}}\right)\hbar . \label{entm3.10}
\end{equation}
This is our improved inequality.  It is a sufficient condition that the state is separable (not entangled).  If it fails, the state must be entangled.  We note that $a^{2}+\frac{1}{a^{2}}$ has a minimum value of 2 at $a=\pm 1$.  Hence

\begin{equation}
\sqrt{\left\langle\left(x_{1}+\frac{a}{|a|}x_{2}\right)^{2}\right\rangle \left\langle\left(p_{1}-\frac{a}{|a|}p_{2}\right)^{2}\right\rangle}\geqslant\hbar . \label{entm3.11}
\end{equation}
This is the necessary condition that a two-particle state be separable.  Thus, we have two necessary conditions, corresponding to choosing $a$ to be positive or negative.  Using (\ref{entm2.4}) and (\ref{entm2.7}), we may write the two conditions in the succinct forms

\begin{equation}
\langle X^{2}\rangle \langle p^{2}\rangle \geqslant\frac{1}{4}\left(\frac{\hbar^{2}}{4}\right), \label{entm3.12}
\end{equation}
and

\begin{equation}
\langle x^{2}\rangle \langle P^{2}\rangle \geqslant 4\left(\frac{\hbar^2}{4}\right). \label{entm3.13}
\end{equation}
We now with to apply these results to the particular state discussed in Sec. II.  Thus, using the results given in (\ref{entm2.13}) and (\ref{entm2.15}), together with (\ref{entm2.9}) and (\ref{entm2.10}), these conditions take the explicit forms

\begin{equation}
d^{2}+\left(\frac{\hbar t}{4md}\right)^{2}\geqslant\frac{\sigma^{2}}{4} \label{entm3.14}
\end{equation}
and

\begin{equation}
\sigma^{2}+
\left(\frac{\hbar t}{m\sigma}\right)^{2}\geqslant 4d^{2}. \label{entm3.15}
\end{equation}
Since we assumed $2d>\sigma$, it follows that (\ref{entm3.14}) (which corresponds to the choice of positive $a$) is automatically fulfilled, implying separability.  However, (\ref{entm3.15}) (which corresponds to the choice of negative $a$) is only fulfilled if

\begin{eqnarray}
t &\geqslant& \left(\frac{m}{\hbar}\right)\left\{\sigma^{2}\left(4d^{2}-\sigma^{2}\right)\right\}^{1/2} \nonumber  \\
&=& \left(\frac{2m\sigma d}{\hbar}\right)~\left(1-\frac{\sigma^{2}}{4d^{2}}\right)^{1/2}\equiv t_{d}. \label{entm3.16}
\end{eqnarray}
Thus, for $t<t_{d}$, the separability condition is violated and the state is entangled.  However at $t=t_{d}$, we encounter ESD [6-8], despite the fact that we know from our exact analysis in Sec. III that the state is entangled for all times. The solution to this apparent contradiction stems from the fact that the Duan et al. condition is a sufficient criterion for inseparability (entanglement) but it is not necessary.  Recognizing this, these authors were led to develop a necessary and sufficient condition for entanglement by using a variety of local linear unitary transformations (consisting of various rotations and squeezing transformations of $x_{1}$ and $p_{1}$ that preserve the commutation relations and similarly for $x_{2}$ and $p_{2}$), and referred to as LOCC, to map any Gaussian state into what they refer to as Standard forms I and II, which eventually leads them to a state for which their separability criterion is both necessary and sufficient.  It is known that these local operations do not affect the entanglement of the state i.e. we have a family of states all with the same entanglement \cite{bennett97,horodecki}.  

Guided by the fact that our results are satisfactory at $t=0$, supplemental by the detailed results which we already obtained for the motion of a free particle \cite{ford02}, especially equation (9) of the latter reference, we make the following local canonical transformations:

\begin{equation}
\bar{x}_{1}(t)=\exp (-\frac{i}{\hbar}H_{1}t)x_{1}(t)\exp (\frac{i}{\hbar}H_{1}t);~\bar{x}_{2}(t)=\exp (-\frac{i}{\hbar}H_{2}t)x_{2}(t)\exp (\frac{i}{\hbar}H_{2}t), \label{entm3.17}
\end{equation}
where 

\begin{equation}
H_{1}=p^{2}_{1}/2m; ~H_{2}=p^{2}_{2}/2m. \label{entm3.18}
\end{equation}
It follows that [since $p_{1}(t)=p_{1}(0)$ and $p_{2}(t)=p_{2}(0)$]

\begin{equation}
\bar{x}_{1}(t)=x_{1}(t)-\frac{t}{m}~ p_{1}(0)=x_{1}(0), \label{entm3.19}
\end{equation}

\begin{equation}
\bar{x}_{2}(t)=x_{2}(t)-\frac{t}{m}~p_{2}(0)=x_{2}(0). \label{entm3.20}
\end{equation}
In fact, the second equality in the latter two equations readily follows from the Heisenberg equation of motion. These transformations lead to the results

\begin{equation}
\bar{x}(t)=x(t)-\frac{t}{m_{x}}~p(0)=x(0), \label{entm3.21}
\end{equation}

\begin{equation}
\bar{X}(t)=X(t)-\frac{t}{M}~P(0)=X(0). \label{entm3.22}
\end{equation}
It follows that the necessary conditions for separability now becomes

\begin{equation}
\langle \bar{X}^{2}\rangle \langle p^{2}\rangle \geq \frac{1}{4} \left(\frac{\hbar^{2}}{4}\right) \label{entm3.23}
\end{equation}
and

\begin{equation}
\langle \bar{x}^{2}\rangle \langle P^{2}\rangle \geq 4\left(\frac{\hbar^{2}}{4}\right) \label{entm3.24}
\end{equation}
which takes the explicit forms

\begin{equation}
2d\geq \sigma \label{entm3.25}
\end{equation}
and

\begin{equation}
\sigma \geq 2d , \label{entm3.26}
\end{equation}
which are only compatible for $\sigma =2d$.  However, since we assumed that $2d>\sigma$, it is clear that the above analysis leads to the conclusion that the system is entangled for all times. We note that the unitary transformations given in (\ref{entm3.17}) led to the elimination of terms depending on $t$ in the separability conditions.

A similar analysis may be carried out using the Peres-Horodecki criterion \cite{peres96,horodecki97}, namely that a state is separable if the partial transpose of the density matrix is a positive operator.  It can be shown explicitly (See Appendix A) that, for our model, it leads to the same result (\ref{entm3.11}), that was obtained from the Duan et al. criterion.

Although the model we are considering here is a pure state, it is instructive to see how it fits into the general framework of {\underline{mixed}} states which are best considered using Wigner distributions \cite{duan00,simon00}.

It should be emphasized at the outset that not all Wigner functions are permissible distribution functions since the corresponding density matrix elements must be positive definite \cite{hillery84} and the uncertainty relations must be satisfied.  We now briefly review the work of Duan et al. \cite{duan00} and Simon \cite{simon00}, which will result in bringing $M$ into the "standard form" \cite{duan00} given in (\ref{entm3.30}) below. 

Recalling that Gaussian states are completely characterized by their first and second moments (and here we have arranged that the former are zero), it follows that the Wigner characteristic function for a Gaussian state of a pair of particles can be written in the general form

\begin{equation}
\tilde{W}(Q_{1},P_{1};Q_{2},P_{2};t)=\exp \left\{ -\frac{\mathbf{Q}\cdot 
\mathbf{M}\cdot \mathbf{Q}}{2}\right\} ,  \label{entm3.27}
\end{equation}
where

\begin{equation}
\mathbf{Q}=\left( 
\begin{array}{c}
\frac{LP_{1}}{\hbar } \\ 
\frac{Q_{1}}{L} \\ 
\frac{LP_{2}}{\hbar } \\ 
\frac{Q_{2}}{L}
\end{array}
\right) ,\qquad \mathbf{M}=\left( 
\begin{array}{cc}
\mathbf{G} & \mathbf{C} \\ 
\mathbf{C}^{T} & \mathbf{H}
\end{array}
\right) .  \label{entm3.28}
\end{equation}
Here $\mathbf{M}$ is the correlation (variance) matrix and $\mathbf{G}$ and $\mathbf{C}$ are $2\times 2$ matrices given by

\begin{eqnarray}
\mathbf{G} &\mathbf{=}&\left( 
\begin{array}{cc}
\frac{\left\langle x_{1}^{2}\right\rangle }{L^{2}} & \frac{\left\langle
x_{1}p_{1}+p_{1}x_{1}\right\rangle }{2\hbar } \\ 
\frac{\left\langle x_{1}p_{1}+p_{1}x_{1}\right\rangle }{2\hbar } & \frac{
L^{2}\left\langle p_{1}^{2}\right\rangle }{\hbar ^{2}}
\end{array}
\right) ,  \notag \\
\mathbf{H} &=&\left( 
\begin{array}{cc}
\frac{\left\langle x_{2}^{2}\right\rangle }{L^{2}} & \frac{\left\langle
x_{2}p_{2}+p_{2}x_{2}\right\rangle }{2\hbar } \\ 
\frac{\left\langle x_{2}p_{2}+p_{2}x_{2}\right\rangle }{2\hbar } & \frac{
L^{2}\left\langle p_{2}^{2}\right\rangle }{\hbar ^{2}}
\end{array}
\right) ,  \notag \\
\mathbf{C} &\mathbf{=}&\left( 
\begin{array}{cc}
\frac{\left\langle x_{1}x_{2}\right\rangle }{L^{2}} & \frac{\left\langle
x_{1}p_{2}\right\rangle }{\hbar } \\ 
\frac{\left\langle x_{2}p_{1}\right\rangle }{\hbar } & \frac{
L^{2}\left\langle p_{1}p_{2}\right\rangle }{\hbar ^{2}}
\end{array}
\right) .  \label{entm3.29}
\end{eqnarray}
In these expressions $L$ and $\hbar$ are constants introduced to make the matrix variance (correlation)$M$
dimensionless. However, as far as the subsequent analysis is concerned, the $L$ may be ignored with impunity, as we will do henceforth.

Making use of a series of local linear canonical transformations (rotations and squeezings), it was shown \cite{duan00,simon00} that it is possible to bring $\mathbf{M}$ to the special form:

\begin{equation}
\mathbf{M}^{\prime }=\left( 
\begin{array}{cccc}
g & 0 & c & 0 \\ 
0 & g & 0 & c^{\prime } \\ 
c & 0 & h & 0 \\ 
0 & c^{\prime } & 0 & h
\end{array}
\right) .  \label{entm3.30}
\end{equation}
Since determinants are invariant under these transformations we have the
following simple relations for determining the quantities $g$, $h$, $c$ and
$ c^{\prime }$, in terms of four invariants,

\begin{eqnarray}
\det \mathbf{G} &=&g^{2},\qquad \det \mathbf{H}=h^{2},  \notag \\
\det \mathbf{C} &=&cc^{\prime },\quad \det \mathbf{M}=\left( gh-c^{2}\right)
\left( gh-c^{\prime 2}\right) .  \label{entm3.31}
\end{eqnarray}
We now turn to the special case of interest here, that is the pure Gaussian state given in (\ref{entm2.1}).  Using the techniques developed in \cite{ford072}, or, since we are dealing with a free particle, from the Wigner function given in (\ref{entmA5}), together with the results given in (\ref{entm2.14}) and (\ref{entm2.15}), it follows that

\begin{eqnarray}
\left\langle x_{1}^{2}\right\rangle  &=&\left\langle x_{2}^{2}\right\rangle
=\left( \frac{1}{\sigma ^{2}}+\frac{1}{4d^{2}}\right) \left[ \sigma
^{2}d^{2}+\left( \frac{\hbar t}{2m}\right) ^{2}\right] ,  \notag \\
\left\langle x_{1}x_{2}\right\rangle  &=&\left( -\frac{1}{\sigma
^{2}}+\frac{ 1}{4d^{2}}\right) \left[ -\sigma ^{2}d^{2}+\left(
\frac{\hbar t}{2m}\right) ^{2}\right] ,  \notag \\
\left\langle p_{1}^{2}\right\rangle  &=&\left\langle p_{2}^{2}\right\rangle
=\left( \frac{1}{2}\right) ^{2}\left( \frac{1}{\sigma ^{2}}+\frac{1}{
4d^{2}}\right) ,  \notag \\
\left\langle p_{1}p_{2}\right\rangle  &=&\left( \frac{1}{2}\right)
^{2}\left( -\frac{1}{\sigma ^{2}}+\frac{1}{4d^{2}}\right) ,  \notag \\
\left\langle x_{2}p_{1}\right\rangle  &=&\left\langle
x_{1}p_{2}\right\rangle =\frac{\hbar ^{2}t}{4m}\left( -\frac{1}{\sigma
^{2}}+
\frac{1}{4d^{2}}\right) ,  \notag \\
\frac{\left\langle x_{1}p_{1}+p_{1}x_{1}\right\rangle }{2} &=&\frac{
\left\langle x_{2}p_{2}+p_{2}x_{2}\right\rangle }{2}=\frac{\hbar
^{2}t}{4m}
\left( \frac{1}{\sigma ^{2}}+\frac{1}{4d^{2}}\right) . \label{entm3.32}
\end{eqnarray}
Hence

\begin{equation}
\mathbf{G}=\mathbf{H} =\left( \frac{1}{\sigma ^{2}}+\frac{1}{4d^{2}}\right)
\left(
\begin{array}{cc}
\sigma ^{2}d^{2}+\left( \frac{\hbar t}{2m}\right) ^{2} & \frac{\hbar
^{2}t}{ 4m} \\
\frac{\hbar ^{2}t}{4m} & \left( \frac{1}{2}\right) ^{2}
\end{array}
\right) \label{entm3.33}
\end{equation}

\begin{equation}
\mathbf{C} =\left( -\frac{1}{\sigma ^{2}}+\frac{1}{4d^{2}}\right)
\left(
\begin{array}{cc}
-\sigma ^{2}d^{2}+\left( \frac{\hbar t}{2m}\right) ^{2} & \frac{\hbar
^{2}t}{ 4m} \\
\frac{\hbar ^{2}t}{4m} & \left( \frac{1}{2}\right) ^{2}
\end{array}
\right). \label{entm3.34}
\end{equation}
It follows that

\begin{equation}
\det \mathbf{G}=\det \mathbf{H}=\left(\frac{1}{\sigma^{2}}+\frac{1}{4d^{2}}\right)^{2}\sigma^{2}d^{2} \label{entm3.35}
\end{equation}

\begin{equation}
\det \mathbf{C}=-\frac{1}{4}\left(-\frac{1}{\sigma^{2}}+\frac{1}{4d^{2}}\right)^{2}\sigma^{2}d^{2} \label{entm3.36}
\end{equation}

\begin{equation}
\det \mathbf{M}=\det (\mathbf{G}+\mathbf{C})\det (\mathbf{G}-\mathbf{C}) . \label{entm3.37}
\end{equation}
We note that the latter two equations are independent of $t$. Thus, the transformed matrix $M$ has the form (\ref{entm3.30}) with

\begin{equation}
g^{2}=\frac{1}{4}\left(\frac{1}{\sigma^{2}}+\frac{1}{4d^{2}}\right)^{2}\sigma^{2}d^{2} \label{entm3.38}
\end{equation}

\begin{equation}
cc^{\prime} = -\frac{1}{4}\left(-\frac{1}{\sigma^{2}}+\frac{1}{4d^{2}}\right)^{2}\sigma^{2}d^{2} \label{entm3.39}
\end{equation}

\begin{equation}
\left(g^{2}-c^{2}\right)\left(g^{2}-c^{\prime 2}\right) = \frac{1}{16}. \label{entm3.40}
\end{equation}
The solution of these equations is

\begin{equation}
g=\frac{1}{2}\sigma d\left(+\frac{1}{\sigma^{2}}+\frac{1}{4d^{2}}\right) \label{entm3.41}
\end{equation}

\begin{equation}
c=-c^{\prime}=-\frac{1}{2}\sigma d\left(-\frac{1}{\sigma^{2}}+\frac{1}{4d^{2}}\right). \label{entm3.42}
\end{equation}
In terms of these quantities, the inequality (\ref{entm3.11}) becomes

\begin{equation}
\sqrt{\left(g\mp c \right)\left(g\pm c^{\prime}\right)}\geqslant 1. \label{entm3.43}
\end{equation}
In terms of the above expressions, this becomes

\begin{equation}
\sigma d\left(\left(\frac{1}{\sigma^{2}}+\frac{1}{4d^{2}}\right)\pm\left(-\frac{1}{\sigma^{2}}+\frac{1}{4d^{2}}\right)\right)\geqslant 1, \label{entm3.44}
\end{equation}
which implies

\begin{equation}
\sigma \geqslant 2d,~~~ \textnormal{or}~~~ 2d\geqslant \sigma . \label{entm3.45}
\end{equation}
In other words, the condition for separability only holds when

\begin{equation}
2d=\sigma \label{entm3.46}
\end{equation}
that is when the center-of-mass coordinate width and the relative coordinate width are equal.  As a consequence,

\begin{equation}
a_{12}=0, \label{entm3.47}
\end{equation}
as expected. As indicated above, these conclusions hold at all times. 

It is also of interest to note that, within the present context, the Peres-Horodecki criterion implies that a Gaussian state is separable if and only if the minimum value of its symplectic spectrum of $M^{T_{2}}$ is greater than $1/2$ \cite{vidal02,simon00} which leads to a good measure of entanglement for all Gaussian states given by

\begin{equation}
E=\max\left\{0,-\log (2\nu_{min})\right\} \label{entm3.48}
\end{equation}
where $\nu_{min}$ is the smallest sympletic eigenvalue of $M^{T_{2}}$.  The equation determining the sympletic spectrum is \cite{vidal02}

\begin{equation}
\nu^{4}+\left(\det \mathbf{G}+\det \mathbf{H}-2\det \mathbf{C}\right)\nu^{2}+\det \mathbf{M}=0, \label{entm3.49}
\end{equation}
with solutions $\pm i\nu_{\alpha}$, $\alpha =1,2$ where $\nu_{\alpha}$ is the symplectric spectrum.  Hence, using (\ref{entm3.35}) and (\ref{entm3.36}), we obtain $\nu_{1}=(d/\sigma )$ and $\nu_{2}=\left(\sigma /4d\right)$. Since we assume $2d>\sigma$, we see that $\nu_{min}=\nu_{2}<\frac{1}{2}$ and hence

\begin{equation}
E=\max\left\{0,\log\left(\frac{2d}{\sigma}\right)\right\}=\log\left(\frac{2d}{\sigma}\right), \label{entm3.50}
\end{equation}
in agreement with the result (\ref{entm4.10}) arising from the $\log$ negativity criterion, as discussed in the next section.

\section{Logarithmic Negativity Criterion}

The logarithmic negativity is defined as

\begin{equation}
E_{N}(\rho )=\log
 \left\{2N(\rho )+1\right\} , \label{entm4.1}
\end{equation}
where $N(\rho )$ is the negativity of the state and is given by the absolute sum of the negative eigenvalues of the partial transpose of $\rho$ \cite{vidal02,plenio05}.

We want to solve the eigenfunction equation:

\begin{equation}
\int dx_{1} \int dx_{2}\langle x^{\prime}_{1},x^{\prime}_{2}\vert\rho^{T_{2}}\vert x_{1},x_{2}\rangle\phi (x_{1},x_{2})=\lambda\phi (x^{\prime}_{1},x^{\prime}_{2}), \label{entm4.2}
\end{equation}
especially in order to obtain the negative eigenvalues. After some algebra (See Appendix B), we find that the eigenvalues are given by

\begin{equation}
\lambda_{mn} =
\begin{cases}
\pm \beta^{m+n} &\lambda_{0}~\text{for}~m\neq n \\
\beta^{2n} &\lambda_{0}~\text{for}~m=n , \label{entm4.3}
\end{cases}
\end{equation}
where $m,n=0, 1, 2, $ -~-~- and $\lambda_{0}$ is a positive eigenvalue given by

\begin{equation}
\lambda_{0}=\frac{2\sqrt{a^{2}_{11}-a^{2}_{12}}}{a_{11}+\sqrt{a^{2}_{11}-a^{2}_{12}}} . \label{entm4.4}
\end{equation}
In addition,

\begin{equation}
\beta = \sqrt{\frac{a_{11}-\sqrt{a^{2}_{11}-a^{2}_{12}}}{a_{11}+\sqrt{a^{2}_{11}-a^{2}_{12}}}} . \label{entm4.5}
\end{equation}
For the symmetric case, these reduce to

\begin{equation}
\lambda_{0}=\frac{8\sigma d}{(2d+\sigma )^{2}} , \label{entm4.6}
\end{equation}
and

\begin{equation}
\beta=\frac{2d-\sigma}{2d+\sigma}, \label{entm4.7}
\end{equation}
recalling that we have assumed that $2d>\sigma$. As a check, we note that

\begin{equation}
\sum_{mn}	\lambda_{mn}=\left(1+\beta^{2}+\beta^{4}+.~.~.\right)\lambda_{0}=\left(\frac{1}{1-\beta^{2}}\right)\lambda_{0}=1, \label{entm4.8}
\end{equation}
verifying that $Tr\rho^{T_{2}}=1$.  In addition,

\begin{equation}
N(\rho )=\sum_{m>n}\vert \lambda_{mn}\vert =\left(\frac{\beta}{1-\beta}\right)=\left(\frac{2d-\sigma}{2\sigma}\right). \label{entm4.9}
\end{equation} 
Hence

\begin{eqnarray}
E_{N}(\rho ) &=& \log\left\{2N(\rho )+1\right\} \nonumber \\
&=& \log\left\{\frac{2d}{\sigma}\right\}. \label{entm4.10}
\end{eqnarray}
Thus, the greater $2d$ is compared to $\sigma$, the larger the negativity and hence the greater the entanglement.  In addition, since $\rho(t)=\exp\left(-iHt\right)\rho (0)\exp\left(iHt\right)$, where $H=\left(p^{2}_{1}+p^{2}_{2}\right)/2m$, it is clear, from (\ref{entm4.2}), that the eigenvalues of $\rho^{T_{2}}$ are invariant under this local unitary transformation. Hence, the result (\ref{entm4.10}) is valid for all times.

\section{Entanglement of Formation}

For bipartite pure states, the entanglement of formation is given by \cite{bennett96}

\begin{equation}
E_{F}=S_{1}(\rho_{1})=S_{2}(\rho_{2}) \label{entm5.1}
\end{equation}
where $\rho_{1}$ and $\rho_{2}$ are the reduced density matrices [defined in (\ref{entmB15})] and

\begin{equation}
S_{1}=-Tr_{2} [\rho_{1}\log\rho_{1}] \label{entm5.2}
\end{equation}
is the von Neuman entropy.  Thus, from appendix B, we have

\begin{equation}
S=-\sum^{\infty}_{n=0}\lambda_{n}\log\lambda_{n}, \label{entm5.3}
\end{equation}
where the eigenvalues of $\rho_{1}$ are given by

\begin{equation}
\lambda_{n}=\lambda_{0}\beta^{2n}~~(n =0, 1, 2, -~-~-), \label{entm5.4}
\end{equation}
where, from (\ref{entm4.6}) and (\ref{entm4.7}), we have (with $R\equiv 2d/\sigma)$

\begin{equation}
\lambda_{0}=\frac{4R}{(1+R)^{2}}=1-\beta^{2} \label{entm5.5}
\end{equation}
and

\begin{equation}
\beta =\frac{R-1}{R+1}. \label{entm5.6}
\end{equation}
Hence

\begin{eqnarray}
S_{1} &=& -\lambda_{0}\sum^{\infty}_{n=0}\beta^{2n}~\left[\log\lambda_{0}+2n\log\beta\right] \nonumber \\
&=& -\lambda_{0}\log\lambda_{0}\sum^{\infty}_{n=0}\beta^{2n}-2\lambda_{0}\log\beta\sum^{\infty}_{n=0}n(\beta^{2})^{n} \nonumber \\
&=& -\log\lambda_{0}-\frac{2\beta^{2}}{1-\beta^{2}}\log\beta \nonumber \\
&=& \log R+\left[2\log\frac{1+R}{2R}+\frac{(R-1)^{2}}{2R}\log\frac{R+1}{R-1}\right]. \label{entm5.7}
\end{eqnarray}

We recall, from (\ref{entm4.10}), that the logarithmic negativity $E_{N}(\rho )$ is given by $\log R$.  Also it can be shown that $0\leq S_{1}\leq \log R$ for $R\geq 1$ and the equalities hold for $R=1$ and $R\rightarrow\infty$, respectively.  This is consistent with the result that the entanglement of formation is always less than logarithmic negativity, and they are equal for maximally entangled pure states.  Thus, the entanglement exists if $2d>\sigma$.

\section{Conclusions}

We examined a simple but non-trivial model of entanglement which enabled us to carry out an \underline{exact} analysis.  We analyzed various entanglement criteria, arising especially from the work of Duan et al. \cite{duan00}; Peres-Horodecki \cite{peres96,horodecki97}; Vidal and Werner \cite{vidal02}, who considered both the logarithmic negativity and that arising from a determination of the smallest sympletic eigenvalue of the Peres transform of the transformed variance matrix and Bennett et al. \cite{chan04} on the entanglement of formation. We found that all of these various entanglement criteria led to the same results but that some are more straightforward than others.  In particular, it was clear that the logarithmic criterion is the simplest to use since the procedure is straightforward, that is obtain the eigenvalues of the Peres transform of the density matrix.

After this paper was completed, we became aware (courtesy of the referee) of various papers that have closely related themes.  Our work is an example of "entanglement without dissipation," which apparently was initially discussed by Chan and Eberly \cite{chan04} who also investigated a Gaussian state but used a Schmidt-state analysis as a measure of entanglement.  Next, Yonac et al. \cite{yonac06} considered two isolated atoms each in their own lossless Jaynes-Cummings cavity.  They showed that, due to the interaction with the local lossless cavities, ESD occurs for atom-atom entanglement due to information loss to the cavity modes but that entanglement is resurrected in a periodic manner following each ESD event due to the fact that the time evolution is lossless.  The same system was analyzed by Sainz and Bjork \cite{sainz07} who concluded that the atoms simply transfer their entanglement to the cavity fields and that an entanglement measure exists that is constant under the time evolution.  A different system, photoionization in a lossless environment, was considered by Fedorov et al. \cite{fedorov04}, who found narrowing of electron and ion wave packets due to electron-ion entanglement.

Entanglement of formation \cite{bennett96} is one of the measures we have discussed (see Sec. V) and this quantity is referred to by Munro et al. \cite{munro01} as "- - the canonical measure of entanglement - -," who then go on to present a class of states that have the maximum amount of entanglement for a given linear entropy.  All of this work is leading to a better understanding but, to quote from the recent general overview of Yu and Eberly \cite{yu09}, "- - there is still no deep understanding of sudden death dynamics."

\section{Acknowledgment}

This work was partially supported by the National Science Foundation under Grant No. ECCS-0757204.

\appendix
\section{Density Matrix Elements and Wigner Functions}

In general, the Peres partial transpose of the density matrix is
\begin{equation}
\left\langle x_{1}^{\prime },x_{2}^{\prime }\left\vert \rho
^{T_{2}}\right\vert x_{1},x_{2}\right\rangle =\left\langle x_{1}^{\prime
},x_{2}\left\vert \rho \right\vert x_{1},x_{2}^{\prime }\right\rangle .
\label{entmA1}
\end{equation}
However, in practice, it is often more convenient to consider the corresponding result for the Wigner function \cite{simon00}, that is

\begin{equation}
\mathcal{W}^{T_{2}}(q_{1},p_{1};q_{2},p_{2})=\mathcal{W}
(q_{1},p_{1};q_{2},-p_{2}).  \label{entmA2}
\end{equation}
The corresponding transpose of the Wigner Characteristic function is

\begin{equation}
\tilde{W}^{T_{2}}(Q_{1},P_{1};Q_{2},P_{2})=\tilde{W}
(Q_{1},P_{1};-Q_{2},P_{2}),  \label{entmA3}
\end{equation}

We recall that the most general Gaussian \underline{pure} state corresponds to the wave function
\begin{equation}
\psi (x_{1},x_{2})=\frac{(a_{11}a_{22}-a_{12}^{2})^{1/4}}{\sqrt{2\pi }}\exp
\left\{ -\frac{a_{11}x_{1}^{2}+2a_{12}x_{1}x_{2}+a_{22}x_{2}^{2}}{4}\right\}
,  \label{entmA4}
\end{equation}
where $a_{11}$ and $a_{22}$ are positive and $a_{11}~a_{22}-a_{12}^{2}>0$ (to ensure integrability). It is straightforward to obtain the Wigner function

\begin{eqnarray}
\mathcal{W}(q_{1},p_{1};q_{2},p_{2}) &=& \frac{1}{(\pi\hbar )^{2}}\exp\left\{-\frac{a_{11}q^{2}_{1}+2a_{12}q_{1}q_{2}+a_{22}q^{2}_{2}}{2}\right\} \nonumber \\
&&\times\exp\left\{-2\frac{a_{22}p^{2}_{1}-2a_{12}p_{1}p_{2}+a_{11}p^{2}_{2}}{\hbar^{2}\left(a_{11}a_{22}-a^{2}_{12}\right)}\right\}, \label{entmA5}
\end{eqnarray}
and the Wigner characteristic function (Fourier transform of the Wigner function)

\begin{eqnarray}
\tilde{W}(Q_{1},P_{1};Q_{2},P_{2}) &=& \exp\left\{-\frac{a_{11}Q^{2}_{1}+2a_{12}Q_{1}Q_{2}+a_{22}Q^{2}_{2}}{8}\right\} \nonumber \\
&& \times\exp\left\{-\frac{a_{22}P^{2}_{1}-2a_{12}P_{1}P_{2}+a_{11}P^{2}_{2}}{2(a_{11}a_{22}-a^{2}_{12})\hbar^{2}}\right\} . \label{entmA6}
\end{eqnarray}
Also, the corresponding density matrix is 

\begin{gather}
\langle x^{\prime}_{1},x^{\prime}_{2} \vert\rho\vert x_{1},x_{2}\rangle = \psi \left(x^{\prime}_{1},x^{\prime}_{2}\right)\psi^{*}(x_{1},x_{2}) \nonumber \\
 =\frac{\left(a_{11}a_{22}-a^{2}_{12}\right)^{1/2}}{2\pi} \nonumber \\
\times\exp\left\{-\frac{a_{11}\left(x^{2}_{1}+x^{\prime 2}_{1}\right)+2a_{12}\left(x_{1}x_{2}+x^{\prime}_{1}x^{\prime}_{2}\right)+a_{22}\left(x^{2}_{2}+x^{\prime 2}_{2}\right)}{4}\right\}. \label{entmA7}
\end{gather}
The Peres partial transpose of this density matrix is

\begin{gather}
\langle x^{\prime}_{1},x^{\prime}_{2}\vert\rho^{T_{2}}\vert x_{1},x_{2}\rangle = 
\psi\left(x^{\prime}_{1},x_{2}\right)\psi^{*} \left(x_{1},x^{\prime}_{2}\right)
=\frac{(a_{11}a_{22}-a^{2}_{12})^{1/2}}{2\pi} \nonumber \\
\times\exp\left\{-\frac{a_{11}(x^{2}_{1}+x^{\prime 2}_{1})+2a_{12}(x_{1}x^{\prime}_{2}+x^{\prime}_{1}x_{2})+a_{22}(x^{2}_{2}+x^{\prime 2}_{2})}{4} \right\} . \label{entmA8}
\end{gather}
The corresponding transpose of the Wigner characteristic function is

\begin{eqnarray}
\tilde{W}^{T_{2}}\left(Q_{1},P_{1};Q_{2},P_{2}\right) &=&\exp \left\{
-\frac{a_{11}Q_{1}^{2}-2a_{12}Q_{1}Q_{2}+a_{22}Q_{2}^{2}}{8}
\right\}   \notag \\
&&\times \exp \{-\frac{a_{22}P_{1}^{2}-2a_{12}P_{1}P_{2}+a_{11}P_{2}^{2}}{
2(a_{11}a_{22}-a_{12}^{2})\hbar ^{2}}\}.  \label{entmA9}
\end{eqnarray}
In addition, the corresponding transpose of the Wigner function is

\begin{gather}
\mathcal{W}^{T_{2}}(q_{1},p_{1};q_{2},p_{2})
=\frac{1}{\left( \pi \hbar \right) ^{2}}\exp \left\{ -\frac{
a_{11}q_{1}^{2}+2a_{12}q_{1}q_{2}+a_{22}q_{2}^{2}}{2}\right\}   \notag \\
\times \exp \left\{
-2\frac{a_{22}p_{1}^{2}+2a_{12}p_{1}p_{2}+a_{11}p_{2}^{2} }{\hbar
^{2}\left( a_{11}a_{22}-a_{12}^{2}\right) }\right\} .  \label{entmA10}
\end{gather}

Consider the symmetric case, for which
\begin{equation}
a_{22}=a_{11}=\frac{1}{\sigma ^{2}}+\frac{1}{4d^{2}},\quad
a_{12}=-\frac{1}{
\sigma ^{2}}+\frac{1}{4d^{2}}.  \label{entmA11}
\end{equation}
Then

\begin{gather}
\langle x^{\prime}_{1},x^{\prime}_{2}\vert\rho^{T_{2}}\vert x_{1},x_{2}\rangle = \frac{(a^{2}_{11}-a^{2}_{12})^{1/2}}{2\pi} \nonumber \\
\exp\left\{-\frac{a_{11}(x^{2}_{1}+x^{\prime 2}_{1})+2a_{12}(x_{1}x^{\prime}_{2}+x^{\prime}_{1}x_{2})+a_{11}(x^{2}_{2}+x^{\prime 2}_{2})}{4} \right\} , \label{entmA12}
\end{gather}

\begin{eqnarray}
\tilde{W}^{T_{2}}(Q_{1},P_{1};Q_{2},P_{2}) &=&\exp \left\{ -\frac{\left(
Q_{1}+Q_{2}\right) ^{2}}{8\sigma ^{2}}-\frac{\left( Q_{1}-Q_{2}\right)
^{2}}{ 32d^{2}}\right\}  \notag \\
&&\times \exp \{-\frac{d^{2}}{2\hbar ^{2}}\left( P_{1}+P_{2}\right) ^{2}-
\frac{\sigma ^{2}}{8\hbar ^{2}}\left( P_{1}-P_{2}\right) ^{2}\},  \label{entmA13}
\end{eqnarray}

\begin{gather}
\mathcal{W}^{T_{2}}(q_{1},p_{1};q_{2},p_{2})=\frac{1}{\left( \pi \hbar
\right) ^{2}}\exp \left\{ -\frac{\left( q_{1}-q_{2}\right) ^{2}}{2\sigma
^{2} }-\frac{\left( q_{1}+q_{2}\right) ^{2}}{8d^{2}}\right\}  \notag \\
\times \exp \left\{ -\frac{2d^{2}}{\hbar ^{2}}\left( p_{1}-p_{2}\right)
^{2}-
\frac{\sigma ^{2}}{2\hbar ^{2}}\left( p_{1}+p_{2}\right) ^{2}\right\}.
\label{entmA14}
\end{gather}

If $\mathcal{W}^{T_{2}}$ is to be a Wigner function we must require that the
uncertainty relation be satisfied. In particular we must require
\begin{equation}
\Delta x\Delta p\geq \frac{\hbar }{2},  \label{entmA15}
\end{equation}
where

\begin{eqnarray}
\left( \Delta x\right) ^{2} &=&\left\langle \left( x_{1}-x_{2}\right)
^{2}\right\rangle ^{T_{2}}  \notag \\
&=&\int dq_{1}\int dp_{1}\int dq_{2}\int dp_{2}\left( q_{1}-q_{2}\right)
^{2}
\mathcal{W}^{T_{2}}(q_{1},p_{1};q_{2},p_{2})  \notag \\
&=&\int dq_{1}\int dp_{1}\int dq_{2}\int dp_{2}\left( q_{1}-q_{2}\right)
^{2}
\mathcal{W}\left( q_{1},p_{1};q_{2},p_{2}\right)  \notag \\
. &=&\left\langle \left( x_{1}-x_{2}\right) ^{2}\right\rangle ,  \notag \\
4\left( \Delta p\right) ^{2} &=&\left\langle \left( p_{1}-p_{2}\right)
^{2}\right\rangle ^{T_{2}}  \notag \\
&=&\int dq_{1}\int dp_{1}\int dq_{2}\int dp_{2}\left( p_{1}-p_{2}\right)
^{2}
\mathcal{W}^{T_{2}}(q_{1},p_{1};q_{2},p_{2})  \notag \\
&=&\int dq_{1}\int dp_{1}\int dq_{2}\int dp_{2}\left( p_{1}+p_{2}\right)
^{2}
\mathcal{W}\left( q_{1},p_{1};q_{2},p_{2}\right)  \notag \\
&=&\left\langle \left( p_{1}+p_{2}\right) ^{2}\right\rangle .  \label{entmA16}
\end{eqnarray}
Therefore, if we require that the Peres transpose corresponds to a Wigner
function, we see from (\ref{entmA15}) and (\ref{entmA16}) that the following inequality must be satisfied

\begin{equation}
\sqrt{\left\langle \left( x_{1}-x_{2}\right) ^{2}\right\rangle \left\langle
\left( p_{1}+p_{2}\right) ^{2}\right\rangle }\geq \hbar .  \label{entmA17}
\end{equation}
This is just the refined condition of Duan et al.

\section{Eigenvalues and Eigenfunctions of the Density Matrix and the Reduced Density Matrix}

For the case of the pure
Gaussian state given by (\ref{entm2.1}), and specializing to the symmetric case, where $
a_{22}=a_{11}$, we found that the Peres transpose is [see (\ref{entmA8})]
\begin{equation}
\left\langle x_{1}^{\prime },x_{2}^{\prime }\left\vert \rho
^{T_{2}}\right\vert x_{1},x_{2}\right\rangle =\psi\left(x^{\prime}_{1},x_{2}\right)\psi^{*}\left(x_{1},x^{\prime}_{2}\right),  \label{entmb1}
\end{equation}
where [see (\ref{entm2.1})]

\begin{equation}
\psi\left(x_{1},x^{\prime}_{2}\right)=\frac{\left(a^{2}_{11}-a^{2}_{12}\right)^{1/4}}{\sqrt{2\pi}}\exp\left\{-\frac{a_{11}\left(x^{2}_{1}+x^{\prime 2}_{2}\right)+2a_{12}x_{1}x_{2}^{\prime}}{4}\right\}, \label{entmB2}
\end{equation}
with a corresponding result for $\psi (x^{\prime}_1,x_{2})$. We want to solve the eigenfunction equation:

\begin{equation}
\int dx_{1}\int dx_{2}\left\langle x_{1}^{\prime },x_{2}^{\prime }\left\vert
\rho ^{T_{2}}\right\vert x_{1},x_{2}\right\rangle \Phi (x_{1},x_{2})=\lambda
\Phi (x_{1}^{\prime },x_{2}^{\prime }).  \label{entmB3}
\end{equation}

The form of (\ref{entmB2}) suggests use of the Mehler formula \cite{merzbacher98}, which is written in terms of the Hermite functions $H_{n}$.  However, we find it is more useful to modify this formula so that it is now written in terms of the related orthogonal function $\phi_{n}$, the eigenstate of the quantum harmonic oscillator.  Thus, for arbitrary variables $x$ and $y$, the modified Mehler formula may be written in the form

\begin{equation}
\sqrt{\frac{\gamma}{\pi}}\exp\left\{\frac{\gamma}{1-\beta^{2}}\left[2\beta xy-\frac{1+\beta^{2}}{2}\left(x^{2}+y^{2}\right)\right]\right\}=\sqrt{1-\beta^{2}}\sum^{\infty}_{n=0}\beta^{n}\phi_{n}(x)\phi_{n}(y), \label{entmB4}
\end{equation}
where the function $\phi_{n}(x)$ is related to the Hermite polynomial $H_{n}(x)$ by

\begin{equation}
\phi_{n}(x)=\sqrt{\frac{1}{2^{n}n!}}\left(\frac{\gamma}{\pi}\right)^{1/4}e^{-\gamma x^{2}/2}H_{n}\left(\sqrt{\gamma}~x\right), \label{entmB5}
\end{equation}
and

\begin{equation}
\int\phi_{m}(x)\phi_{n}(x)dx=\delta_{mn}. \label{entmB6}
\end{equation}
Thus, applying (\ref{entmB4}) to both $\psi(x_{1},x^{\prime}_{2})$ and $\psi(x^{\prime}_{1},x_{2})$, we obtain

\begin{equation}
\langle x^{\prime}_{1},x^{\prime}_{2}\vert\rho^{T_{2}}\vert x_{1},x_{2}\rangle =\lambda_{0}\sum_{m,n}\beta^{m+n}\phi_{m}(x_{1})\phi_{m}(x^{\prime}_{2})\phi_{n}(x^{\prime}_{1})\phi_{n}(x_{2}). \label{entmB7}
\end{equation}
Here the parameters $\gamma$, $\beta$, and $\gamma_{0}$ are given by

\begin{equation}
\gamma = \sqrt{a^{2}_{11}-a^{2}_{12}}/2 \label{entmB8}
\end{equation}

\begin{equation}
\beta=\sqrt{\frac{a_{11}-\sqrt{a^{2}_{11}-a^{2}_{12}}}{a_{11}+\sqrt{a^{2}_{11}-a^{2}_{12}}}} \label{entmB9}
\end{equation}

\begin{equation}
\lambda_{0}=\frac{2\sqrt{a^{2}_{11}-a^{2}_{12}}}{a_{11}+\sqrt{a^{2}_{11}-a^{2}_{12}}}=1-\beta^{2}. \label{entmB10}
\end{equation}
The advantage of writing the matrix elements in terms of orthogonal functions, as given in (\ref{entmB7}), immediately leads to solutions of the eigenfunction equation (\ref{entmB3}), in the form

\begin{equation}
\int dx_{1}\int dx_{2}\langle x^{\prime}_{1},x^{\prime}_{2}\vert\rho^{T_{2}}\vert x_{1},x_{2}\rangle~\phi_{m}(x_{1})\phi_{n}(x_{2})=\lambda_{0}\beta^{m+n}\phi_{m}(x^{\prime}_{1})\phi_{n}(x^{\prime}_{2}). \label{entmB11}
\end{equation}
If $m=n$, $\Phi_{nn}(x_{1},x_{2})=\phi_{n}(x_{1})\phi_{n}(x_{2})$ is the desired eigenfunction with eigenvalue $\lambda_{nn}=\beta^{2n}\lambda_{0}$.  On the other hand if $m\neq n$, we find that the eigenfunction is given by

\begin{equation}
\Phi_{mn}\left(x_{1},x_{2}\right)=\frac{1}{\sqrt{2}}\left[\phi_{m}(x_{1})\phi_{n}(x_{2})\pm\phi_{n}(x_{1})\phi_{m}(x_{2})\right] \label{entmB12}
\end{equation}
with eigenvalues $\pm\lambda_{0}\beta^{m+n}$.  In summary, the complete eigenvalues are

\begin{equation}
\lambda_{mn} =
\begin{cases}
\pm \beta^{m+n}\lambda_{0}~~~&\text{for}~~m\neq n \\
~~~\beta^{2n}\lambda_{0}~~~&\text{for}~~m=n , 
\end{cases} \label{entmB13}
\end{equation}
with $m,n=0,1,.~.~.$

Next, we consider reduced density matrices.  Thus, we define

\begin{equation}
\rho_{1}=Tr_{2}\langle x^{\prime}_{1},x^{\prime}_{2}\vert\rho^{T_{2}}\vert x_{1},x_{2}\rangle . \label{entmB14}
\end{equation}
This is obtained by setting $x_{2}=x^{\prime}_{2}$ in (\ref{entmB7}) and integrating over $x_{2}$ to get

\begin{equation}
\rho_{1}=\lambda_{0}\sum_{m}\beta^{2m}\phi_{m}(x_{1})\phi_{m}(x^{\prime}_{1}). \label{entmB15}
\end{equation}
Thus, the eigenvalue equation for the reduced density matrix is

\begin{equation}
\int dx_{1}\rho_{1}\Phi^{(1)}(x_{1})=\lambda\Phi^{(1)}(x^{\prime}_{1}) .\label{entmB16}
\end{equation}
It immediately follows that this leads to solutions $\phi_{m}(x_{1})$ with corresponding eigenvalues $\lambda_{0}\beta^{2m}(m=0, 1, 2 -~-~-)$.  Moreover, it is clear that the corresponding eigenvalues of $\rho_{2}$ are the same.

\end{document}